 \def\cm#1{}

\def\comment#1{}
\documentstyle[pra,epsf,aps]{revtex}
\begin{document}
\title{Theory and Satellite Experiment
for Critical Exponent $ \alpha $ \\ of $ \lambda $-Transition in
Superfluid Helium
}
\author{Hagen Kleinert%
 \thanks{Email: kleinert@physik.fu-berlin.de ~~ URL:
http://www.physik.fu-berlin.de/\~{}kleinert ~~ Phone/Fax:
 0049 30 8383034 }}
\address{Institut f\"ur Theoretische Physik,\\
Freie Universit\"at Berlin, Arnimallee 14,
14195 Berlin, Germany}
%\pacs{03.20.+i\\ 04.20.Fy\\ 02.40.+m}
%%%%%%%%%%%%
\maketitle
\begin{abstract}
On the basis recent seven-loop perturbation expansion
for $ \nu^{-1} =3/(2- \alpha )$
we perform a careful reinvestigation
of the critical exponent
$ \alpha $
governing the
 power behavior $|T_c-T|^{- \alpha }$
of the specific heat of superfluid helium near the phase transition.
With the help of variational strong-coupling theory.
we find
$ \alpha=-0.01126\pm0.0010$,
in very good agreement with the
space shuttle experimental value $ \alpha =-0.01056\pm0.00038$.
\end{abstract}

%\pacs{03.20.+i\\ 04.20.Fy\\ 02.40.+m}
%%%%%%%%%%%%
%\maketitle
%
\noindent
{\bf 1.} The critical exponent
$ \alpha $ characterizing the
 power behavior $|T_c-T|^{- \alpha }$
of the specific heat of superfluid helium
near the transition temperature $T_c$
is presently the best-measured critical exponent of all.
A microgravity
experiment in the Space Shuttle in October 1992
rendered a value
with amazing
 precision  \cite{rLipa}
\begin{equation}
 \alpha^{\rm ss} =-0.01056\pm0.00038.
\label{@spaceshuttal}\end{equation}
This represents a considerable
change and improvement of
the
experimental number
 found a long time ago on earth
by G. Ahlers \cite{ahl}:
\begin{equation}
 \alpha =-0.026\pm0.004 ,
\label{@ahlers}\end{equation}
in which the sharp peak of the specific heat
was broadened to $10^{-6}\,$K by the tiny pressure difference
between top and bottom of the sample.
In space, the temperature could be
brought to within $10^{-8}\,$K
close to $T_c$ without seeing this broadening.

The exponent $ \alpha $ is extremely sensitive
to the precise value of the critical exponent $ \nu $
 which determines the growth
of the coherence length
when approaching  the critical
temperature, $\xi\propto |T-T_c|^{- \nu }$.
Since $ \nu $ lies very close to
$2/3$, and $ \alpha $ is related to $ \nu $ by the scaling relation
$ \alpha =2-3 \nu $, a tiny change of $ \nu $
produces a large relative change of $ \alpha $.
Ahlers' value was for many years an embarrassment
to quantum field theorists
who never could find $ \alpha $ quite as negative ---
the field theoretic $  \nu  $-value came usually out smaller
than  $ \nu _{\rm Ahl}=0.6753\pm0.0013$.
The space shuttle measurement was therefore extremely welcome
since
it comes much closer to previous theoretical values.
In fact, it turned out to agree extremely well with the most recent
theoretical determination of $ \alpha $ by
strong-coupling perturbation theory \cite{kl}
based on the recent seven-loop
power series expansions
of $ \nu $ \cite{MN,nickel,anton},
which gave
\cite{seven}
\begin{equation}
 \alpha^{\rm sc} =-0.0129\pm0.0006.
\label{@earlierr}\end{equation}

The purpose of this Letter
is to present yet another resummation
of
the
perturbation expansion for $ \nu^{-1}$
and for $ \alpha =2-3 \nu $
by variational perturbation theory,
applied in a different way
than in \cite{seven}.
Since it is a priori unclear which of the two results
should be more accurate,
we combine them
to the slightly less negative average value with a larger
error
\begin{equation}
 \alpha^{\rm sc} =-0.01126\pm0.0010.
\label{@crexpo}\end{equation}

Before entering the more technical
part of the paper, a few comments are necessary
on the reliability
of error estimates for any theoretical result of this kind.
They can certainly be trusted no more than the experimental numbers.
Great care went into the analysis of Ahlers` data \cite{ahl}.
Still, his final result (\ref{@ahlers})
does not accommodate the space shuttle value
(\ref{@spaceshuttal}). The same surprise may happen
to theoretical
results and their error limits in papers on resummation of divergent
perturbation expansions,
since there exists so far no safe way of
determining the errors.
The expansions in powers of the coupling constant $g$ are strongly divergent,
and one knows
accurately only the first
seven coefficients, plus the leading
growth behavior for large orders $k$ like
$ \gamma   (-a)^k k! k   \Gamma(k+b)$. The parameter
$b$ is determined by the number of zero modes in
a solution  to a classical field equation,
$a$ is the inverse energy of this solution,
and $ \gamma $ the entropy of its small oscillations.

The shortness of the available expansions and their divergence
make  estimates of the error range of the result a
rather subjective procedure.
All publications resumming critical exponents such as $ \alpha $
calculate
some sequences of
$N$th-order resummed  approximations
$ \alpha _N$,
and estimate an error range from
the way these
tend to their limiting value.
While these estimates may be statistically significant,
there are unknown systematic errors.
Otherwise one should be able to take the expansion for any
function $\tilde f(g)\equiv f( \alpha(g) )$
and find a limiting number
$f( \alpha )$ which lies in the
corresponding range of values.
This is unfortunately not true in general.
Such reexpansions
can approach their limiting values in many different ways,
and it is not clear which yields the most reliable result.
One must therefore seek as much additional information on the series
as possible.

One such additional information
becomes available by
resumming the expansions in powers of the bare coupling constant $g_0$
rather than the renormalized one $g$.
The reason is that
any function of the bare coupling constant
 $f(g_0)$
which has a finite critical limit
approaches this limit
with a nonleading inverse power of $g_0^ \omega $, where $ \omega $
is called the {\em critical exponent of approach to scaling\/},
whose size is known to be about $0.8$ for superfluid helium.
Any resummation method which naturally incorporates
his power behavior should converge faster than those
which ignore it.
This incorporation is precisely
the virtue of
variational perturbation theory, which we have therefore
chosen for the resummation of $ \alpha $.

For a second additional information
we take advantage of our theoretical knowledge
on the general form
of the large-order behavior
of the expansion coefficients
\begin{eqnarray} \label{@omas}
 \gamma   (-a)^k k! k   \Gamma(k+b)
\left(1+\frac{c ^{(1)}}{k}        +\frac{c^{(2)}}{k^2}+\dots\right).
\label{@grow}\end{eqnarray}
In the previous paper
\cite{seven}
we have done so by
choosing the nonleading parameters
$c^{i}$
to reproduce exactly the first seven known expansion  coefficients
of $ \alpha $.
The resulting  expression
(\ref{@grow}) determines all expansion coefficients.
The so-determined  expression
(\ref{@grow}) predicts approximately {\em all\/} expansion coefficients,
with increasing precision for increasing orders.
The extended power series
has then been resummed for increasing orders $N$, and
from the $N$-behavior
we have found the
$ \alpha $-value (\ref{@earlierr}) with quite a small
error range.

As a third additional information we use
the fact that we know from theory \cite{kl}
in which way the infinite-order result is approached.
Thus we may fit the approximate values
$ \alpha _N$ by an appropriate expansion in $1/N$
and achieve in this way a more accurate
estimate of the limiting value than without such an extrapolation.
The error can thus be  made much smaller than the
distance between the last two approximations,
as has been verified in many model studies of divergent series
\cite{JanKl}.

The strategy of this paper goes as follows:
We want to use all the additional informations
on the expansion of the critical exponent $ \alpha $
as above, but apply the variational resummation
method in two more alternative ways.
First, we  reexpand the series $ \alpha (g_0)$ in powers of
a variable
$h$
whose critical limit is no longer infinity but $h=1$.
The closer distance to the expansion point $h=0$
leads us to  expect a faster convergence.
Second we resum  two different expansions,
one for $ \alpha $,  and one for $f( \alpha )= \nu ^{-1}\equiv 3/(2- \alpha )$.
From the difference in the
resulting $ \alpha $-values
and a comparison
 with the earlier result (\ref{@earlierr})
we obtain an estimate of the systematic errors
which is
specified in
Eq.~(\ref{@crexpo}).
~\\

\noindent
{\bf 2.}
The seven-loop power series expansion for $ \nu $ in powers
of the unrenormalized coupling constant
of O(2)-invariant $\phi^4$-theory which lies in the
 universality class of superfluid helium
reads
 \cite{MN}
\begin{eqnarray}
 \nu^{-1}&=&2 - 0.4\,{g_0} + 0.4681481481482289\,{{{g_0}}^2} -
  0.66739\,{{{g_0}}^3} + 1.079261838589703\,{{{g_0}}^4} -
  1.91274\,{{{g_0}}^5}\nonumber \\& +& 3.644347291527398\,{{{g_0}}^6} -
  7.37808\,{{{g_0}}^7}  +\dots~.
\label{@seriesnum}\end{eqnarray}
By fitting the expansion coefficients
with the theoretical large-order behavior (\ref{@grow}),
this
series has been extended to higher orders as follows \cite{seven}
\begin{eqnarray}
 \Delta  \nu^{-1}&=&
 15.75313406543747\,{{{g_0}}^8} -
  35.2944\,{{{g_0}}^9} + 82.6900901520064\,{{{g_0}}^{10}} -
  202.094\,{{{g_0}}^{11}} +
  514.3394395526179\,{{{g_0}}^{12}}\nonumber \\& -&
  1361.42\,{{{g_0}}^{13}} +
  3744.242656157152\,{{{g_0}}^{14}} -
  10691.7\,{{{g_0}}^{15}}               +\dots~.
\label{@seriesnumD}\end{eqnarray}

The renormalized coupling constant is related to the unrenormalized one by
an expansion $g=\sum _{k=1}^7a_kg_0^k$.
Its power behavior for large $g_0$ is determined by a
series
\begin{eqnarray}
s=\frac{d\log g(g_0)}{d\log g_0} &=&
1 - {g_0} + {\frac{947\,{{{g_0}}^2}}{675}} -
  2.322324349407407\,{{{g_0}}^3} +
  4.276203609026057\,{{{g_0}}^4}\nonumber \\ &-&
  8.51611440473227\,{{{g_0}}^5} +
  18.05897631325589\,{{{g_0}}^6}+\dots~.
\label{@sdirect}\end{eqnarray}
A similar best fit
of these by the theoretical large-order behavior extends this series
by
\begin{eqnarray}
  \Delta s&=&
  40.38657228730114\,{{{g_0}}^7} +
  94.6453399123477\,{{{g_0}}^8} -
  231.3922442162566\,{{{g_0}}^9} +
  588.3206172579102\,{{{g_0}}^{10}}\nonumber \\ &-&
  1552.116358404217\,{{{g_0}}^{11}} +
  4242.372685080157\,{{{g_0}}^{12}} -
  12001.18866491822\,{{{g_0}}^{13}} +
  35115.23006646194\,{{{g_0}}^{14}}\nonumber \\ &-&
  106234.4643086436\,{{{g_0}}^{15}} +
  332239.2175082959\,{{{g_0}}^{16}} +\dots     ~.
\label{@delsext}\end{eqnarray}
Scaling implies that $g(g_0)$ becomes
a constant for $g_0\rightarrow \infty$,
implying that the power $s$ goes to  zero in this limit.
By inverting the expansion for $s$, we obtain an expansion
for $ \nu ^{-1}$ in powers of $h\equiv 1-s$ as follows:
\begin{eqnarray}
 \nu ^{-1}(h)&=&2 \!-\! 0.4\,{h} \!-\! 0.093037\,{{{h}}^2} \!+\!
  0.000485012\,{{{h}}^3} \!-\! 0.0139286\,{{{h}}^4} \!+\!
  0.007349\,{{{h}}^5} \!-\! 0.0140478\,{{{h}}^6} \!+\!
  0.0159545\,{{{h}}^7} \!-\! 0.029175\,{{{h}}^8}\nonumber \\ &+&
  0.0521537\,{{{h}}^9} \!-\! 0.102226\,{{{h}}^{10}} \!+\!
  0.224026\,{{{h}}^{11}} \!-\! 0.491045\,{{{h}}^{12}} \!+\!
  1.22506\,{{{h}}^{13}} \!-\! 3.00608\,{{{h}}^{14}} \!+\!
  8.29528\,{{{h}}^{15}} \!-\! 22.5967\,{{{h}}^{16}} .
\label{@seriesdirfornu}\end{eqnarray}
This series has to be evaluated at $h=1$.
For estimating the systematic errors of our resummation,
we also calculate from (\ref{@seriesdirfornu})
a series for $ \alpha =2-3 \nu $
\begin{eqnarray}
 \alpha (h)&=&0.5 \!-\! 0.3\,{h} \!-\! 0.129778\,{{{h}}^2} \!-\!
  0.0395474\,{{{h}}^3} \!-\! 0.0243203\,{{{h}}^4} \!-\!
  0.0032498\,{{{h}}^5} \!-\! 0.0121091\,{{{h}}^6}\nonumber \\&+&
  0.00749308\,{{{h}}^7} \!-\! 0.0194876\,{{{h}}^8}\!+\!
  0.0320172\,{{{h}}^9} \!-\! 0.0651726\,{{{h}}^{10}} \!+\!
  0.14422\,{{{h}}^{11}} \!-\! 0.315055\,{{{h}}^{12}}
\nonumber \\&+&
  0.802395\,{{{h}}^{13}} \!-\! 1.95455\,{{{h}}^{14}} \!+\!
  5.49143\,{{{h}}^{15}} \!-\! 14.8771\,{{{h}}^{16}} \!+\! \dots~.
\label{@seriesdirforal}\end{eqnarray}

\begin{figure}[tbhp]
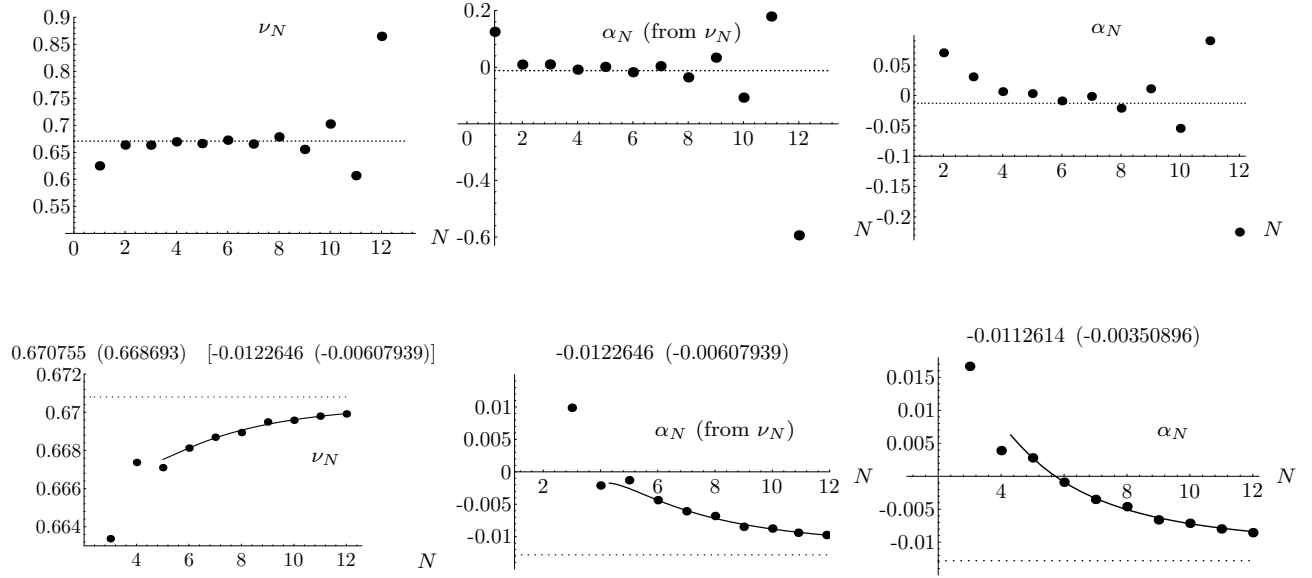

\input nual0.tps ~\\
\input nualr.tps
\caption[]{Upper plots:
Results of partial sums of
series   (\protect\ref{@seriesdirfornu})
for $ \nu ^{-1}$ up to order $N$, once plotted as $ \nu _N=1/ \nu _N^{-1}$,
and once as $ \alpha _N=2-3 \nu _N$. The third plot shows the
corresponding partial sums of
the series for $ \alpha $. The dotted line is the experimental
space shuttle value $ \alpha^{\rm ss}$ of Eq.~(\ref{@spaceshuttal}).
Lower plots:
The corresponding resummed values and
a fit of them by
$c_0+c_1/N^2+c_2/N^4$.
The constant $c_0$ is written on top, together with
the seventh-order approximation (in parentheses).
The square brackets on top of the left-hand plot for $ \nu $
shows the corresponding $ \alpha $-values.
}
\label{@partialsums}\end{figure}

{}~\\
\noindent
{\bf 3.}
In order to get a rough
idea about the behavior
of the reexpansions
in powers of $h$,
 we
plot their partial sums at $h=1$
in the upper row of Fig.~\ref{@partialsums}.
After an initial apparent convergence, these
show the typical divergence of
perturbation expansions.

A rough resummation  is possible using Pad\'e approximants.
The results are shown in Table~\ref{@padeal}.
The highest Pad\'e approximants yield
\begin{equation}
 \alpha ^{\rm Pad}=-0.0123\pm0.0050.
\label{@}\end{equation}
The error is estimated by the distance to the next lower approximation.

\begin{figure}[tbhp]
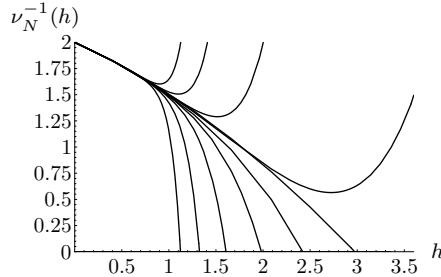

~~~~~~~~~~~~~~~~~~~~~~~~~~~~~~~~\input nuipl.tps
\caption[]{Successive truncated expansions of $ \nu ^{-1}(h)$
of orders $N=2,\dots,12$.}
\label{fignuipl}\end{figure}

{}~\\
\noindent
{\bf 4.}
We now
resum the expansions
$ \nu ^{-1}(h)$ and $ \alpha (h)$
by variational perturbation theory.
This is applicable to
divergent perturbation expansions
\begin{equation}
f(x)  =   \sum_{n=0}^{\infty} a_n
      x ^n  ,
\label{wcexpnew}\end{equation}
which behave for large $x$ like
\begin{equation}
f(x)  =  x^{p/q} \sum_{m=0}^{\infty} b_m
      x ^{-2m/q}
\label{wcexpnewsc}\end{equation}
It is easy to adapt our function to this general
behavior. Plotting the successive truncated
power series for $ \nu ^{-1}(h)$ against $h$ in Fig.~\ref{fignuipl},
we see that this function
will have a zero somewhere above $h=h_0=3$.
We therefore go over to the variable $x$ defined by
$h=h(x)\equiv h_0x/(h_0-1+x)$, in terms of which  $f(x)= \nu ^{-1}(h(x))$
behaves like (\ref{wcexpnewsc}) with $p=0$ and $q=2$,
and has to be evaluated at $x=1$.
The large-$x$ behavior is imposed upon
the function with the expansion (\ref{wcexpnew}) as follows.
We insert
an auxiliary  scale parameter $ \kappa $ and
define the truncated  functions
\begin{equation}
f_N(x)  \equiv  \kappa ^p   \sum_{n=0}^{N} a_n
     \left(\frac{ x  }{ \kappa^q }\right)^n .
\label{wcexp}\end{equation}
The parameter $ \kappa $
will be
 set
 equal to $1$ at the end.
Then we introduce a variational parameter $K$ by the replacement
\begin{equation}
 \kappa \rightarrow \sqrt{K^2+ \kappa ^2-K^2}.
\label{@paramrepl}\end{equation}
The functions $f_N(x)$ are so far independent of $K$.
This is changed by expanding
the square root in (\ref{@paramrepl})
in powers of $ \kappa ^2-K^2$,
thereby treating this difference as a quantity of order
$x$.
This transforms the  terms $ \kappa ^px^n/ \kappa ^{qn}$ in
(\ref{wcexp}) into  polynomials of
$r\equiv ( \kappa ^2-K^2)/K^2$:
\begin{equation}
\kappa ^p \frac{x^n}{ \kappa ^{qn}}\rightarrow K^p\frac{x^n}{K^{qn}}\left[ 1+
\left((p-qn)/2\atop~~1\right)r
+\left((p-qn)/2\atop~~2\right)r^2+\dots
+\left((p-qn)/2\atop N-n\right)r^{N-n}\right] ,
\label{@binexp}\end{equation}
Setting now $ \kappa =1$, and replacing the variational parameter
$K$ by  $v $ defined by
$K^2\equiv x/v$, we obtain from
(\ref{wcexp}) at $x=1$ the variational expansions
\begin{equation}
f_N(v)  =
 \sum_{n=0}^{N} a_n v^{qn-p/2}
\left[1+(v-1)\right]^{(p-qn)/2}_{N-n},
\label{wcexpvar}\end{equation}
where the symbol $\left[1+A\right]^{(p-qn)/2}_{N-n}$ is a short
notation for the
binomial expansion
of $(1+A)^{(p-qn)/2}$ in powers of $A$ up to the order $A^{N-n}$.

\begin{figure}[t]
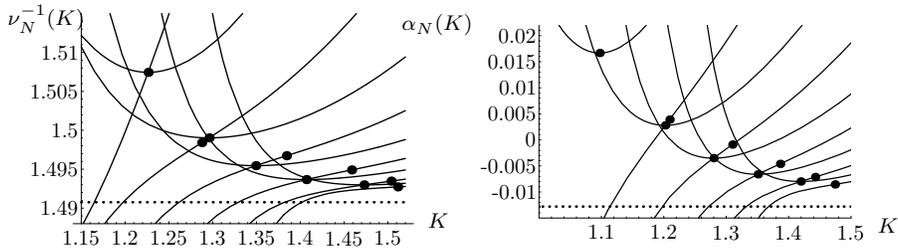

\input albehn2.tps
\caption[]{Successive variational functions
$ \nu ^{-1}_N(h)$ and $ \alpha _N(h)$
with $N=3,\dots,12$  of Table~\protect\ref{@expal}
plotted for $h=x=1$ against the variational parameter $K=\sqrt{x/v}$, together
with their minima for odd $N$, or turning points for even $N$.
These points are plotted against $N$ in the lower row of Fig.~\ref{@partialsums},
where they are extrapolated to $N\rightarrow \infty$,  yielding the critical
exponents.
}
\label{@albehn2}\end{figure}

The variational expansions are optimized in $v$
by minima for odd, and by turning points  for even
$N$, as shown in Fig.~\ref{@albehn2}.
The extrema are plotted as a function of the order $N$ in the lower row
of Fig.~\ref{@partialsums}.
The left-hand plot shows directly the
extremal values of $ \nu_N ^{-1}(v)$, the middle plot shows
the $ \alpha $-values $ \alpha _N=2-3 \nu_N $
corresponding to these. The
right-hand plot, finally, shows the extremal values of $ \alpha _N(v)$.
All three sequences of approximations are fitted very well
by
a large $N$  expansion $c_0+c_1/N^2+c_2/N^4$,
if we omit the lowest five data points which are not yet very regular.
The inverse powers $2$ and $4$ of $N$ in this
fit are determined
by starting from a more general ansatz  $c_0+c_1/N^{p_1}+c_2/N^{p_2}$
and varying $p_1,p_2$ until the sum of the square deviations
of the fit from the points is minimal.
The highest-order data point is taken to be
the one with $N=12$ since,
up to this order, the successive asymptotic
values $c_0$ change monotonously by decreasing amounts. Starting with
$N=13$, the changes increase and reverse direction.
In addition, the mean square deviations of the fits
increasing drastically,
indicating a decreasing usefulness
of the extrapolated expansion coefficients in (\ref{@seriesnumD})
and (\ref{@delsext}) for the extrapolation $N\rightarrow \infty$.
From the parameter $c_0$ of the best fit
for $ \alpha $ which is indicated on top of the
lower right-hand plot in Fig.~\ref{@partialsums},
we find the
critical exponent $ \alpha =-0.01126$ stated in Eq.~(\ref{@crexpo}),
where the error estimate takes into account
the basic systematic errors
indicated by the difference between
the
resummation of $ \alpha =2-3 \nu $,
and of $ \nu ^{-1}$, which by the lower middle plot in Fig.~\ref{@partialsums}
yields $ \alpha =-0.01226$.
It also accommodates
our
earlier
seven-loop strong-coupling result
(\ref{@earlierr})
of Ref.~\cite{seven}.
The dependence on the choice of $h_0$ is negligible
as long as the resummed series $\nu ^{-1}(x)$ and $ \alpha (x)$
do not change their Borel character. Thus $h_0=2.2$ leads to results
well within the
error limits in (\ref{@crexpo}).

\comment{It is interesting to see that
the Pad\'e values (\ref{@Paderes})
carry a much larger error bar
than our individual variational resummation results,
but if one takes into account the
differences between these, the resulting
error estimates
are of the same order of magnitude.}

\begin{figure}[tbhp]
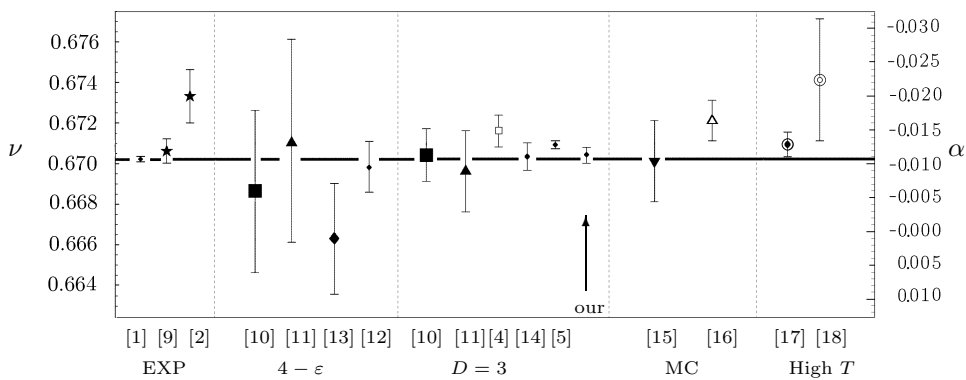

\input nu2my.tps
\caption[]{Survey of experimental and theoretical values for $ \alpha $.
The latter come from resummed perturbation expansions
of $\phi^4$-theory in $4- \varepsilon $ dimensions, in three
dimensions, and
 from high-temperature expansions of XY-models on a lattice.
The sources are indicated below.}
\label{@nu2my}\end{figure}

Our number as well as many  earlier results
are displayed in Fig.~\ref{@nu2my}.
The entire subject is discussed in detail in
the textbook H. Kleinert and V. Schulte-Frohlinde,
{\em Critical Exponents from Five-Loop Strong-Coupling $\phi^4$-Theory
in 4-$ \varepsilon $ Dimensions\/},
  World Scientific, Singapore, 2000
      (http://www.physik.fu-berlin.de/\~{}kleinert/re.html\#b8)

~\\
Acknowledgment\\
The author is grateful to
Dr. J.A. Lipa for several interesting informations
on his experiment.

~\\
Note added in proof:\\
A recent calculation of $ \alpha $ by an improved high-temperature expansion yields the exponent
$ \alpha =-0.0150(17)$ [M. Campostrini, A. Pelissetto, P. Rossi,
and E. Vicari, Phys. Rev. B {\bf 61}, 5905 (2000)].

\begin{table}[tbhp]
\caption[]{Results of the Pad\'e approximations $P_{MN}(h)$ at $h=1$ to  the power series
$ \nu ^{-1}(h)$ and  $ \alpha (h)$. The parentheses show the
associated values of $ \alpha $ and $ \nu $.}
\begin{tabular}{|l|l|l|}
$MN$&$~~~~~~~\nu ~~~~~~~( \alpha )$&$~~~~~~( \nu ) ~~~~~~~~~~\alpha$\\
\hline
{}~4~4& 0.678793 (-0.0363802)~~~~~~~~&(0.678793) -0.0363802~~~~~~~~\\
{}~5~4& 0.671104 (-0.0133107)~~~~~~~~&(0.670965) -0.0128940~~~~~~~~\\
{}~4~5& 0.670965 (-0.0128940)~~~~~~~~&(0.670901) -0.0127031~~~~~~~~\\
{}~5~5& 0.670756 (-0.0122678)~~~~~~~~&(0.670756) -0.0122678~~~~~~~~
\end{tabular}
\label{@padeal}
\label{@Paderes}\end{table}
\scriptsize{
\begin{table}[tbh]
\caption[]{Variational reexpansions of $ \nu _N^{-1}(h)$
and    $  \alpha _N(h)$
for $N=2,\dots,9$
at $h=x=1$ which are plotted in Fig.~\ref{@albehn2}
and whose  minima and turning points
are extrapolated to $N=\infty$ in the lower left- and right-hand plots
 of Fig.~\ref{@partialsums}.
The lists are carried only to $N=9$, to save space, whereas the plots are
for $N=3,\dots,12$.
}
\begin{tabular}{l}
\hline
$\nu^{-1}_2=  2 - 1.2v + 0.69067v^2 $\\
      $\nu^{-1} _3=2 - 1.8v + 2.07200v^2 - 0.72036v^3
$\\$\nu^{-1}_    4=2 - 2.4v + 4.14400v^2 - 2.88145v^3 + 0.53412v^4
$\\$\nu^{-1}_    5=2 - 3.0v + 6.90667v^2 - 7.20363v^3 + 2.67060v^4 +0.28949v^5
$\\$ \nu^{-1}_   6=2 - 3.6v + 10.3600v^2 - 14.4073v^3 +
                               8.01180v^4 + 1.73692v^5 - 2.96286v^6
$\\$ \nu^{-1}_   7=2 - 4.2v + 14.5040v^2 - 25.2127v^3 + 18.6942v^4 +
                               6.07922v^5 - 20.7401v^6 + 11.1835v^7
$\\$\nu^{-1}_    8=2 - 4.8v + 19.3387v^2 - 40.3403v^3 + 37.3884v^4 +
                               16.2113v^5 - 82.9602v^6 + 89.4683v^7 - 36.9575v^8
$\\$\nu^{-1}_    9=2 - 5.4v + 24.8640v^2 - 60.5105v^3 + 67.2992v^4 +
                               36.4753v^5 - 248.881v^6 + 402.607v^7 - 332.617v^8 +
                               121.914v^9
\comment{$\\$ \nu^{-1}_{10}=2 - 6.0v +  31.0800v^2 - 86.4436v^3 +
                               112.165v^4 + 72.9506v^5 - 622.202v^6 + 1342.02v^7 -
                               1663.09v^8 + 1219.14v^9 - 419.968v^{10}
$\\$
   \nu^{-1}_ { 11}=2 - 6.6v + 37.9867v^2 - 118.860v^3 + 176.260v^4 +
                               133.743v^5 - 1368.84v^6 + 3690.57v^7 - 6097.98v^8 +
                               6705.28v^9 - 4619.65v^{10} + 1535.47v^{11}
$\\$\nu^{-1}_{ 12}=2 - 7.2v + 45.5840v^2 - 158.480v^3 + 264.390v^4 +
                               229.273v^5 - 2737.69v^6 + 8857.36v^7 - 18293.9v^8 +
                               26821.1v^9 - 27717.9v^{10}+ 18425.7v^{11} -
                               5974.23v^{12} }$
%\end{tabular}
%\label{@expnum1}\end{table}
%
%\vspace{-.5cm}
%\begin{table}[tbh]
%\caption[]{Variational reexpansions of $  \alpha _N(h)$
%for $N=2,\dots,9$
%at $h=1$ which are plotted in Fig.~\ref{@albehn2}
%and whose  minima and turning points
%are extrapolated to $N=\infty$ in the lower right plot
% of Fig.~\ref{@partialsums}.
%The list is carried only up to $N=9$ to save space. The plots are
%for $N=3,\dots,12$.
%}
%\begin{tabular}{l}
~\\
\hline
$\alpha_   2= 0.5 - 0.90 v + 0.3830 v^2 $\\
$\alpha_    3= 0.5 - 1.35
        v + 1.1490 v^2 - 0.26997 v^3 $\\
$\alpha_    4= 0.5 - 1.80
        v + 2.2980 v^2 - 1.07989 v^3 + 0.025254 v^4 $\\
$\alpha_          5= 0.5 - 2.25 v + 3.8300 v^2 - 2.69972 v^3 + 0.126271
        v^4 + 0.57604 v^5 $\\
$\alpha_   6= 0.5 - 2.70 v + 5.7450
        v^2 - 5.39945 v^3 + 0.378812 v^4 + 3.45629 v^5 - 2.19244
        v^6 $\\
$\alpha_   7= 0.5 - 3.15 v + 8.0430 v^2 - 9.44903
        v^3 + 0.883895 v^4 + 12.0970 v^5 - 15.3471 v^6 + 6.89011
        v^7 $\\
$\alpha_  8= 0.5 - 3.60 v + 10.724 v^2 - 15.1184
        v^3 + 1.767790 v^4 + 32.2587 v^5 - 61.3884 v^6 + 55.1208
        v^7 - 21.5704 v^8 $\\
$\alpha_   9= 0.5 - 4.05 v + 13.788
        v^2 - 22.6777 v^3 + 3.182020 v^4 + 72.5821 v^5 - 184.165
        v^6 + 248.044 v^7 - 194.134 v^8 + 70.781 v^9 $
\comment{$\alpha_    {10}= 0.5 - 4.50 v + 17.235 v^2 - 32.3967 v^3 + 5.30337
        v^4 + 145.164 v^5 - 460.413 v^6 + 826.813 v^7 - 970.67
        v^8 + 707.81 v^9 - 247.254 v^{10} $\\
$\alpha_ {11}= 0.5 - 4.95
        v + 21.065 v^2 - 44.5454 v^3 + 8.33386 v^4 + 266.134
        v^5 - 1012.91 v^6 + 2273.73 v^7 - 3559.12 v^8 + 3892.95
        v^9 - 2719.79 v^{10} + 921.578 v^{11} $\\
$\alpha_ {         12}= 0.5 - 5.4 v + 25.278 v^2 - 59.3939 v^3 + 12.5008
        v^4 + 456.23 v^5 - 2025.82 v^6 + 5456.96 v^7 - 10677.4
        v^8 + 15571.8 v^9 - 16318.8 v^{10} + 11058.9 v^{11} - 3652.2
        v^{12} }
\end{tabular}
\label{@expal}
%\label{@expnum1}
\end{table}
}

\end{document}